\documentclass[12pt,preprint]{aastex}
\shorttitle{Pulsar Radio Emission and Polarization} 
\shortauthors{}
\received{}         
\begin{document}   
\title{RELATIVISTIC MODEL ON PULSAR RADIO EMISSION AND POLARIZATION}
\author{D. Kumar$^1$ and R. T. Gangadhara$^2$}  
\affil{Indian Institute of Astrophysics, Bangalore - 5600034, India\\
$^1$dinesh@iiap.res.in, $^2$ganga@iiap.res.in}  
\altaffiltext{}{}     
\begin{abstract} 
We have developed a relativistic model for pulsar radio emission and
polarization by taking into account of detailed geometry of emission
region, rotation and modulation. The sparks activity on the polar cap
leads to plasma columns in the emission region and modulated
emission.  By considering relativistic plasma bunches streaming out
along the rotating dipolar field lines as source of curvature
radiation, deduced the polarization state of the radiation field in
terms of the Stokes parameters. We have simulated a set of typical
pulse profiles, and analyzed the role of viewing geometry, rotation
and modulation on the pulsar polarization profiles.  Our simulations
explain most of the diverse behaviors of polarization generally found
in pulsar radio profiles.  We show that both the `antisymmetric' and
`symmetric' types of circular polarization are possible within the
frame work of curvature radiation.  We also show that the `kinky'
nature in the polarization position angle traverses might be due to
the rotation and modulation effects. The phase lag of polarization
position angle inflection point relative to the phase of core peak
also depends up on the rotationally induced asymmetry in the curvature
of source trajectory and modulation.
\end{abstract}    
\section{INTRODUCTION}
Even though pulsars were discovered more than four decades ago, their
radio emission process is still not completely understood. The very
high degree of linear polarization and systematic polarization
position angle (PPA) swing of pulsar radiation have been naturally
invoked in curvature radiation \citep{S71, RS75}.  In the frame work
of curvature radiation models the radio emission is believed to be emitted
by relativistic plasma streaming `force-freely' along the open field
lines of the super-strong magnetic field, the geometry of which is
assumed to be predominantly dipolar.  In the non-rotating
approximation, the velocity of relativistic plasma will be parallel to
the tangents of the field lines to which they are associated with, and
hence emitted radiation beamed in the direction of field line
tangents. In the rotating vector model (RVM), as the pulsar
rotates, observer sight line encounters different dipolar field
lines, and results in the `S' shaped PPA swing, which is more or less
determined by the geometry of emission region \citep{RC69,K70}.
 
Pulsar radio emission is believed to be coming from mainly open
dipolar field lines which lie within the polar cap region. The shapes
of individual pulses indicate that the entire polar cap might not be
radiating uniformly.  The sub-pulse modulation can be explained on the
idea of isolated sparks on the polar cap \citep[e.g.,][]{RS75,CR80}.
Pulsar average profiles resulted from the summation of several
hundreds of individual pulses, have well defined shapes and in general
they made up of many components.  Phenomenologically, pulsar emission
is recognized as central `core' emission arising from the region near
to the magnetic pole and `cone' emission arising from concentric rings
around the pole \citep[e.g.,][]{R83,R90,R93,MD99,GG01,MR02}. However,
there are some contrary arguments that the emission is `patchy'
\citep{LM88}.

Since pulsars are fast spinning objects, the rotation effects such as
aberration, retardation and polar cap currents are believed to be
strongly influencing their emission. For an inertial observer, in
addition to intrinsic velocity along the field line tangents,
particles will have co-rotation velocity component. Therefore, in the
inertial observer frame, the net velocity of particles will be offset
from the field line tangents to which they are associated with, and
hence the emission will be aberrated in the direction of pulsar
rotation. By taking into account of rotation, Blaskiewicz, Cordes and
Wasserman (1991), hereafter BCW (1991), have proposed a relativistic
pulsar polarization model. By assuming a constant emission altitude
$r$ across the whole pulse, they have predicted that the midpoint of
the intensity profile shifts to the earlier phase by $\sim r/r_{LC}$
with respect to fiducial phase, whereas PPA inflection point shifts to
later phase by $\sim 3 r/r_{LC}.$ The parameter $r_{LC}=c~P/2 \pi$ is
the light cylinder radius, and $c$ is the velocity of light and $P$ is
the pulsar rotation period. Further, they have shown that the
intensity on leading side becomes stronger than that on trailing side
due to rotation, which has strong observational support
\citep{LM88}. \citet{HA01} further improved the relativistic RVM model
by taking into account of induced magnetic field due to polar cap
currents. \citet{D08} has presented a more simple derivation in which
he has reproduced the \citet{BCW91} prediction. Following these
deductions \citet{TG10} have estimated the absolute emission altitudes
of a few pulsars.  The asymmetry in the phase location of conal
components with respect to core has been interpreted in terms of
rotation effects such as aberration and retardation phase shifts
\citep[]{GG01,GG03,DRH04,G05,Ketal09}.

By solving the equation of motion, \citet{TG07} have also predicted
that the emissions on leading side dominate over the trailing side due
to rotation induced asymmetry in the curvature of trajectory of
bunches.  \cite{DWD10} have analyzed the influence of rotation on
shape of pulse profiles of millisecond pulsars and identified two
opposing effects of corotation: (1) the caustic enhancement of the
trailing side emission due to squeezing into a narrower component; and
(2) the weakening of the trailing side caused by the the smaller
curvature of source trajectories. \citet{Tetal10} have discussed the
significance of geometric and rotation effects on the pulsar radio
profiles and interpreted the asymmetry in the pulsar radio profiles,
particularly ``partial cones'' \citep[which were first termed
  by][]{LM88} in terms of the rotation effects. These are notable for
their highly asymmetric average intensity profiles and PPA traverses,
wherein one side of a double component conal profiles is either
missing or significantly suppressed, and the PPA inflection point lies
well towards the trailing side.

Among the several theoretical problems related to exploring the pulsar
radio emission mechanism, the mostly unexplained observational fact is
the high degree of circular polarization and its very diverse
behavior.  By analyzing average pulsar profiles, \citet{RR90} have
identified two types of circular polarization, namely,
`antisymmetric', where the circular polarization changes its sense
near the center of the pulse profile, and `symmetric', where the
circular polarization will have the same sense across the whole pulse
profile.  In the case of pulsars with antisymmetric type, they found a
strong correlation between the sense reversal of circular polarization
and the PPA swing, and speculated that it could be a geometric
property of emission mechanism. \citet{Hanetal98} have noticed that
the circular polarization is common in pulsars but diverse in nature,
and even though it is generally strongest in the central or `core'
regions, is by no means confined to central regions.  They found a
strong correlation between the sense of circular polarization and the
PPA swing in the double-conal pulsars, and no correlation between the
sense reversal of circular polarization near the center of pulse
profiles and the PPA swing in the pulsars with antisymmetric type of
circular polarization.  Further, \citet{YH06} have reconfirmed these
investigations with a larger data.

There are two probable origins of circular polarization proposed for
the pulsar radiation: either intrinsic to the emission mechanism
\citep[e.g.,][]{M87,GS90a,GS90b,RR90,G97,G10} or generated by the
propagation effects \citep[e.g.,][]{CR79,M03}. \citet{Getal93} have
modeled the single-pulse polarization characteristics of pulsar
radiation and argued that the strong sense reversing circular
polarization is a natural feature of curvature radiation. \citet{CR79}
have suggested that conversion of linear polarization to circular
polarization is possible due to expected asymmetry between the
positive and negative charged components of magneto active plasma
in the far magnetosphere. On the other hand, \citet{KMM91} have
argued that the cyclotron instability, rather than propagation effect,
is responsible for the circular polarization of pulsar radiation.  By
considering the rotation of magnetosphere, \citet{LP99} have shown
that the induced wave mode coupling in the polarization-limiting
region can result in circular polarization in linearly polarized
normal waves. \citet{M03} reviewed the properties of intrinsic
circular polarization, and the circular polarization due to cyclotron
instability, and discussed the circular polarization due to
propagation effects in an inhomogeneous birefringent plasma.

Recently, \cite{G10} has developed a curvature radiation model by
considering the detailed geometry of emission region in the
non-rotating pulsar approximation. His results supports the fact that
the circular polarization survives only when there is a sufficient
gradient in the sub-pulse modulation, and the antisymmetric circular
polarization is an intrinsic nature of the curvature radiation. He has
confirmed the \cite{RR90} correlation between the sense reversal of
circular polarization and the PPA swing. He has also shown that the
sense reversal of circular polarization is by no means confined to
central core regions.

Although an extensive pulsar polarimetric studies are available, the
pulsar emission and polarization are not well understood due to their
diverse nature. Despite the rotation effects such as aberration and
retardation are strongly believed to influence the pulsar radio
profiles, a complete polarization model including rotation effects has
not been attempted in the literature. In the work of Thomas and
co-authors, and \cite{DWD10} modeled only the total intensity, and
left the polarization part untouched. On the other hand, relativistic
models proposed by \citet{BCW91}, \citet{HA01}, and \citet{D08} deals
with PPA only, and not even the linear polarization. Further, in their
models, emission from the points at which bunch velocity exactly
aligns with the sight line is only considered, and the emissions from
the neighboring points at which velocity lies within $\sim 1/\gamma$
with respect to sight line are not considered. But such emissions does
have influences on PPA swing if modulated as we show in our model.

For the first time, we have developed a complete polarization model by
taking into account of a detailed geometry of emission region,
rotation effects and modulation. We adopt the features of curvature
radiation, and incorporated the Gaussian sub-pulse modulation. At any
instant of time, observer tends to receive the incoherent curvature
radiation from a modulated beaming region, which constitutes a small
flux tube of dipolar field lines.  Based on simulated profiles, we
discuss the combined effect of rotation and modulation on the typical
pulsar radio profiles.  We ascribe the asymmetry in the pulsar radio
profiles in terms of combined effects of viewing geometry, rotation,
and modulation.  In \S~2 we derive the expressions for radiation
electric field in frequency domain and the Stokes parameters.  In \S~3
we present the simulation of typical pulse profiles, and the
discussion in \S~4 and conclusion in \S~5.

\section{POLARIZATION STATE OF THE RADIATION FIELD}
Relativistic plasma streaming `force-freely' along the super-strong
dipolar magnetic field lines emit beamed curvature radiation. The
curvature radiation model requires an efficient plasma bunching to
account for the very high brightness temperature of the pulsar radio
emission, wherein the plasma bunches of size less than or equal to the
radiation wavelength can emit coherently
\citep[e.g.,][]{S71,RS75,CR80}. In our model, we treat the plasma
bunch as a single particle of charge $q=Ne,$ where $e$ is the
electronic charge, and $N$ is the number of particles. In this paper
we alternatively use source, particle, or plasma bunch as source of
radio emission but they all mean the same.  The emissions from different
bunches become incoherent as such emissions do not bear any phase relation.

\begin{figure}
\centering
\epsscale{.9}
\plotone{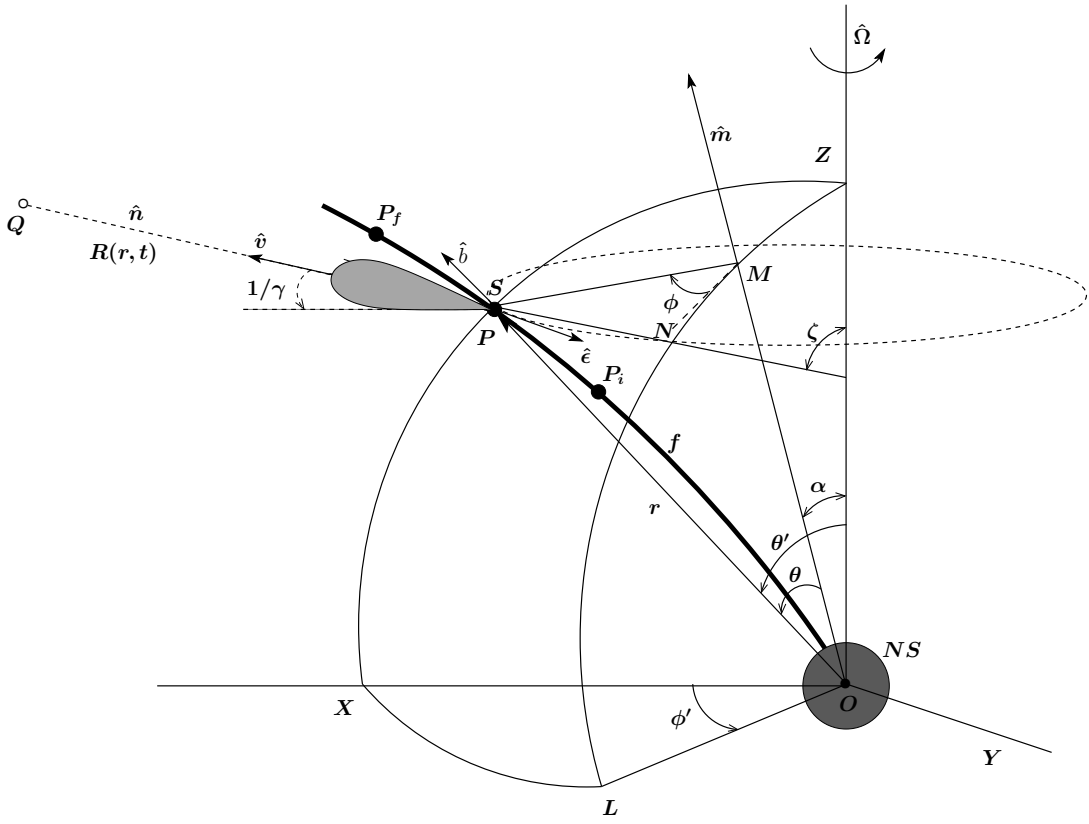}
\caption{Geometry of radiation emission from a relativistic source S
  accelerated along a rotating field line f (thick curve) in a
  stationary inertial frame XYZ with neutron star center O as the
  origin. $\hat{\Omega}$ is the rotation axis, $\hat{m}$ is the
  magnetic axis, and $\hat{b}$ is the field line tangent. The rotation
  direction is $\hat{\epsilon}$ and the net velocity is $\hat{v}$. The
  observer's sight line $\hat{n}$ lies in the fiducial plane
  (XZ-plane). ZPX, ZML, and XL are the great circles centered on
  O. The observation point Q is at a distance R from the emission
  point P.}
 \label{fig:beam_geometry}
\end{figure}

Consider an inclined and rotating magnetic dipole in the inertial
observer's frame (IOF), a stationary Cartesian coordinate system--XYZ
with neutron star center O as the origin as shown in Figure
\ref{fig:beam_geometry}. The angular velocity
$\mbox{\boldmath$\Omega$}$ is considered along positive Z-axis, and
the magnetic axis $\hat{m}$ is inclined by an angle $\alpha$ with
respect to $\mbox{\boldmath$\Omega$}$. Consider a radiation source S
constrained to move along the rotating field line f. The velocity
$\textbf{v}$ of the source is given by
\begin{eqnarray}
\textbf{v} =  \kappa c \,\hat{b} + \mathbf\Omega\times\textbf{r}~,
\label{eqn:v}
\end{eqnarray}
where $\hat{b}$ is unit tangent vector to the field line and
$\textbf{r}$ is the position vector of the source, and their
expressions are given in \citet{G10}. The parameter $\kappa$ specifies
the speed of the source along the field line as a fraction of the
speed of light c. The first term on the r.h.s. of Equation
(\ref{eqn:v}) is the velocity in the corotating frame and is in the
direction of the associated field line tangent, and the second term is
corotation velocity. Hence the velocity of the source is offset from
the field line tangent to which it is associated with, and is aberrated
in the direction of pulsar rotation. The parameter $\kappa$ can be
deduced from the Equation (\ref{eqn:v}) by assuming
$|\textbf{v}|=\beta c$,
\begin{equation} 
\kappa = \sqrt{\beta^{2}-\left(\frac{\Omega r}{c}\right)^{2}
  \sin^{2}\theta'\sin^{2}\Theta} - \frac{\Omega r}{c}
\sin\theta'\cos\Theta~,
\label{eqn:kappa}
\end{equation}
where $\beta=\sqrt{1-1/\gamma^{2}}$, $\gamma$ is the Lorentz factor of
the source, $\theta'$ is the angle between $\textbf{r}$ and
$\mbox{\boldmath$\Omega$}$, and $\Theta$ is the angle between
$\textbf{b}$ and the rotation direction $\hat{\epsilon}$. The
expressions for $\theta'$ and $\Theta$ are given in \citet{G05}.

The acceleration $\textbf{a} = d\textbf{v}/dt$ of source in IOF is
given by
\begin{eqnarray} 
\textbf{a} =\frac{(\kappa
  c)^{2}}{|\textbf{b}|}\frac{\partial\hat{b}}{\partial\theta}+
\frac{\kappa
  c^{2}}{|\textbf{b}|}\frac{\partial\kappa}{\partial\theta}\hat{b}+ 2
\kappa c (\mbox{\boldmath $\Omega$}\times\hat{b}) + \mbox{\boldmath
  $\Omega$}\times(\mbox{\boldmath $\Omega$}\times\textbf{r})~,
\label{eqn:a}
\end{eqnarray}
where we have used the expression of arc length of the field line $ds
= |\textbf{b}| d\theta = \kappa c ~dt$ wherein $|\textbf{b}|$ is the
magnitude of the field line tangent and $\theta$ is magnetic
colatitude. The expression for $|\textbf{b}|$ is given in \cite{G04}.

The first term on r.h.s. of Equation (\ref{eqn:a}) is the acceleration
of bunch due to curvature of dipolar magnetic field line, and this is
the only term which exists in the absence of rotation \citep{G10}. The
second term is due to a small change in the speed of bunch due to
motion along field line. The third and last terms are the
accelerations due to Coriolis and the Centrifugal forces,
respectively.

As the relativistic source accelerates along the rotating field line,
it emits curvature radiation whose spectral distribution at the
observation point Q is given by \citep{G10}
\begin{equation}
\textbf{E}(\textbf{r},\omega) = \frac{1}{\sqrt{2 \pi}}\frac{qe^{i \omega
    R_{0}/c}}{R_{0}~c}\int^{+\infty}_{-\infty}
\frac{\hat{n}\times[(\hat{n} - \mbox{\boldmath
      $\beta$})\times\mbox{\boldmath $\dot{\beta}$}]}{\xi^{2}} e^{i
  \omega ( t-\hat{n}\cdot \textbf{r}/c)} dt~,
\label{eqn:Ew1}
\end{equation}
where $\hat{n}=\left\lbrace \sin\zeta,~0,~\cos\zeta\right\rbrace$ is
the observer's sight line and $\zeta=\alpha + \sigma$ with $\sigma$
being the sight line impact angle. The parameters
$\mbox{\boldmath$\beta$}=\textbf{v}/c$ and $\mbox{\boldmath
  $\dot{\beta}$}=\textbf{a}/c$ are the velocity and acceleration of
source, respectively. $\xi = 1-\mbox{\boldmath$\beta$}\cdot\hat{n},$
and $R_{0}$ is the distance from the neutron star center to 
observer.

At any rotation phase $\phi'=\phi'_{m}$ of the magnetic axis and for a
given emission altitude $r,$ the viewing geometry allows observer to
receive the beamed emission only from a specific region of the pulsar
magnetosphere. Observer receives the maximum radiation from the
emission point at which $\hat{v}$ and $\hat{n}$ are exactly aligned,
and we define the corresponding magnetic colatitude
$\theta=\theta_{0}$ and azimuth $\phi=\phi_{0}.$ By considering a
slowly rotating (or non-rotating) magnetosphere \citet{G04} has
derived the expressions for $\theta_{0}$ and $\phi_{0}$ as functions
of rotation phase $\phi'=\phi'_{m}$ of the magnetic axis \citep[see
Eqns.~9 \& 11 in][]{G04}. We solve $\hat{n}\cdot\hat{v}=1$
numerically to find $\theta_{0}$ and $\phi_{0},$ as the exact
analytical solutions become complicated once the effect of rotation is
considered. Our numerical algorithm starts with $\theta_{0}$ and
$\phi_{0}$ obtained from the \citet{G04} as the initial guess values,
and find the refined ones by solving $\hat{n}\cdot\hat{v}=1.$

\begin{figure}
\centering
\epsscale{.8}
\plotone{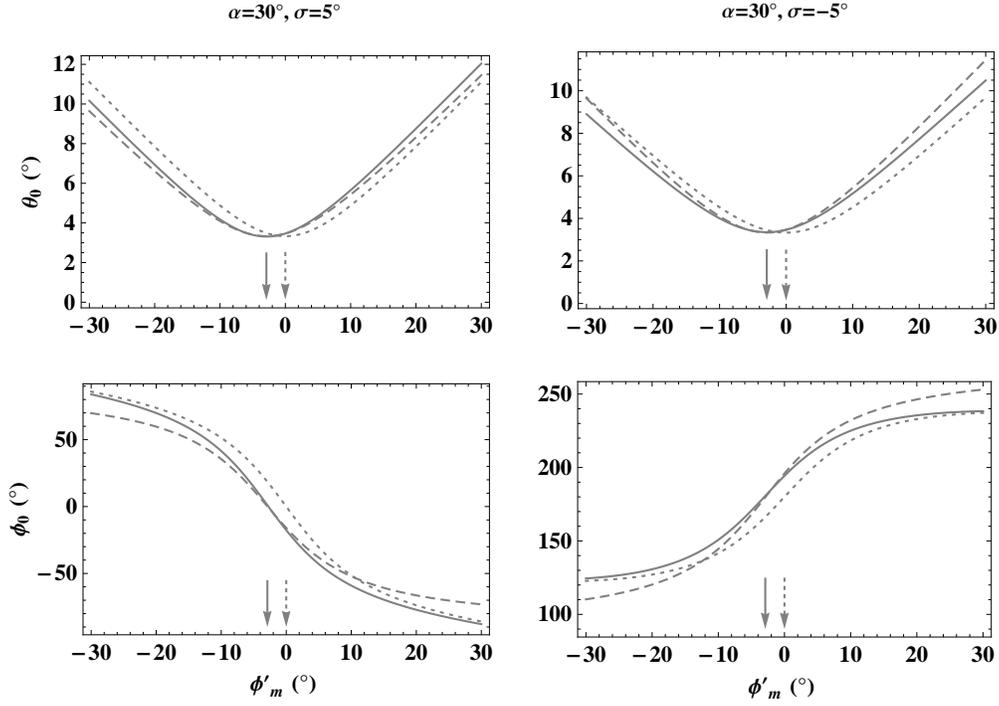}
\caption{The magnetic colatitude $\theta_{0}$ and azimuth $\phi_{0}$
  of the emission point as functions of rotation phase $\phi'_{m}$ in
  the non-rotating (dotted curves) and the rotating (solid 
  curves) cases. The dashed curves are those due to BCW (1991)
  model.  Chosen $P=1$~s, and $r_{n}=r/r_{LC}=0.05.$ The dotted and
  solid arrows in the upper panels represent the location of
  $\theta_{0}$ minima while in the lower panels the
  location of inflection point of $\phi_{0}$ in non-rotating and
  rotating cases, respectively.}
 \label{fig:thetaphi}
\end{figure}

The magnetic colatitude $\theta_{0}$ and azimuth $\phi_{0}$ as
functions of rotation phase $\phi'_{m}$ are plotted in Figure
\ref{fig:thetaphi} for both non-rotating (dotted curves) and rotating
(solid curves) cases by using the parameters $\alpha=30^{\circ}$,
$P=1$~s, $r_{n}=r/r_{LC}=0.05$ and $\sigma=\pm 5^{\circ}$ In the
rotating case both the minimum of $\theta_{0}$ and inflection point of
$\phi_{0}$ shift to the earlier rotation phase compared to those in
non-rotating case.  The absolute phase shifts of both $\theta_{0}$
minimum and the inflection point of $\phi_{0}$ from the fiducial phase
$(\phi'_{m}=0)$ are found to be $2.9^{\circ},$ which is about $\sim
r_n,$ as predicted by \citet{BCW91}. For comparison we have superposed
the curves (dashed line) due to \citet{BCW91} model, and find that
their approximated analytical solutions are valid only over a smaller
rotation phase around $2.9^{\circ}.$

Even though the observer receives maximum radiation from the emission
point $(\theta_{0},\,\phi_{0})$ at which $\hat{v}$ and $\hat{n}$ are
exactly aligned, observer receives a considerable radiation from the
neighboring emission points too due to the finite width of emission
beam. It is about $\sim2/\gamma,$ and the boundary of the emission
region centered on $\hat{n}$ is specified by the condition
$\hat{n}\cdot\hat{v}=\cos(1/\gamma).$ For the computational purpose we
discretized the beaming region into `beaming region points' (BRP), and
the coordinates $\theta$ and $\phi$ of each point is specified by
$\theta_{e}$ and $\phi_{e},$ respectively.  Since the coordinates
$\theta$ and $\phi$ are orthogonal, their ranges
$\theta_{e,min}\leq\theta\leq\theta_{e,max}$ and
$\phi_{e,min}\leq\phi\leq\phi_{e,max}$ can be used to specify the
beaming region boundary, where the subscripts (e,min) and (e,max)
denote the lower and upper boundaries of the emission region. By
considering $\theta_{0}-1/\gamma$ and $\theta_{0}+1/\gamma$ as initial
guess values for $\theta_{e,min}$ and $\theta_{e,max}$ at
$\phi=\phi_{0},$ we solve $\hat{n}\cdot\hat{v}=\cos(1/\gamma)$
numerically and find the roots $\theta_{e,min}$ and $\theta_{e,max}.$
Next for any $\theta=\theta_{e}$ within the range between
$\theta_{e,min}$ and $\theta_{e,max}$, we consider $\phi_{0}-1/\gamma$
and $\phi_{0}+1/\gamma$ as the initial guess values for $\phi_{e,min}$
and $\phi_{e,max}$, and solve again
$\hat{n}\cdot\hat{v}=\cos(1/\gamma)$ numerically to find the roots
$\phi_{e,min}$ and $\phi_{e,max}.$ Note that beaming regions on
leading side become broader compared to those on corresponding
trailing ones, but, they are symmetric in the non-rotating case
\citep{G10}. This is because the source trajectories on leading side
get squeezed whereas on trailing side they get stretched
\citep[e.g.,][]{TG07}.

Note that Equation (\ref{eqn:Ew1}) is the integration of the electric
field of radiation emitted by the relativistic source along its
trajectory. Observer receives the beamed radiation only for a small
segment of source trajectory, say between the points $P_{i}$ and
$P_{f}$ along a rotating field line (see Figure
\ref{fig:beam_geometry}). As the source moves from $P_{i}$ to $P_{f}$
along the rotating field line, time changes from $t_{i}$ to $t_{f}$,
colatitude $\theta$ changes from $\theta_{i}$ to $\theta_{f}$, and the
rotation phase $\phi'$ of magnetic axis changes from $\phi'_{i}$ to
$\phi'_{f}$. From the relation $ds = |\textbf{b}| d\theta = \kappa c
~dt$, we deduce the expression for time $t:$
\begin{equation}
 t=\int \frac{|\textbf{b}|}{\kappa c} d\theta +K~,
\label{eqn:time1}
\end{equation}
where $K$ is the integration constant. Using the condition
$t=\phi'_{m}/\Omega$ for all $\theta=\theta_{e}$ within the
beaming region, it follows from Equation~(\ref{eqn:time1}) that
$K=(\phi'_{m}/\Omega)-\left({\int {|\textbf{b}|/(\kappa c)}
  d\theta}\right)_{\rm \theta=\theta_{e}}$. Therefore we have
\begin{equation}
t=\frac{\phi'_{m}}{\Omega}+\int \frac{|\textbf{b}|}{\kappa c} d\theta
- \left( {\int \frac{|\textbf{b}|}{\kappa c} d\theta}\right) _{\rm
  \theta=\theta_{e}}
\label{eqn:time}
\end{equation}
and the corresponding $\phi'=\Omega t$ for the source motion along any
given field line within the beaming region. Hence the argument of the
integrand in Equation~(\ref{eqn:Ew1}) becomes function of $\theta$
only:
 \begin{equation}
\textbf{E}(\textbf{r},\omega) = \frac{1}{\sqrt{2 \pi}}\frac{qe^{i \omega
    R_{0}/c}}{R_{0}~c}\int^{+\infty}_{-\infty}
\frac{|\textbf{b}|}{\kappa c}\frac{\hat{n}\times[(\hat{n} -
    \mbox{\boldmath $\beta$})\times\mbox{\boldmath
      $\dot{\beta}$}]}{\xi^{2}} e^{i \omega \left\lbrace
  t-\hat{n}\cdot \textbf{r}/c\right\rbrace } d\theta~,
\label{eqn:Ew2}
\end{equation} 
where time $t$ has to be replaced by the expression given in Equation
(\ref{eqn:time}).

Let  
\begin{equation}
\textbf{A} = \{A_{i}\} =\frac{|\textbf{b}|}{\kappa
  c}\frac{\hat{n}\times[(\hat{n} - \mbox{\boldmath
      $\beta$})\times\mbox{\boldmath $\dot{\beta}$}]}{\xi^{2}}~,
\end{equation}
where $A_{i}$ with $i=x,~y$ and $z$, are the components of
$\textbf{A}$ in the $X,$ $Y$ and $Z$ directions, respectively (see
Fig.~\ref{fig:beam_geometry}).  We series expand the components of
$\textbf{A}$ in powers of $\theta$ about $\theta_{e}$, and obtain
\begin{equation}
A_{i}(\theta)=a_{i0}+a_{i1}(\theta-\theta_{e})+a_{i2}(\theta-\theta_{e})^{2}+
O[(\theta-\theta_{e})^{3}]~,
\label{eqn:A}
\end{equation}
where $a_{ij}$ with $i=x,~y$ and $z$, and $j=0,~1$ and $2$ are the
series expansion coefficients. They are given by
\begin{equation}
 a_{i0} = A_{i}(\theta_{e})~,\quad
 a_{i1} = A'_{i}(\theta_{e})~,\quad
 a_{i2} = \frac{1}{2}A''_{i}(\theta_{e})~,
\end{equation}
where $A'_{i}$ and $A''_{i}$ are the respective first and second
derivatives of $A_{i}$ with respect to $\theta$ evaluated at
$\theta_{e}$. Since the expressions of $a_{i0},$ $a_{i1},$ and
$a_{i2}$ are too big, we have not reproduced them here. However one
can always reproduce them by differentiating $\bf A.$

We set the argument of exponential in Equation (\ref{eqn:Ew2}), $\omega
\left\lbrace t-\hat{n}\cdot \textbf{r}/c\right\rbrace = C,$ and series
expand in powers of $\theta$ about $\theta_{e}$, and obtain
\begin{equation}
C (\theta) = c_{0}+
c_{1}(\theta-\theta_{e})+c_{2}(\theta-\theta_{e})^{2}+c_{3}
(\theta-\theta_{e})^{3}+O[(\theta-\theta_{e})^{4}]~,
\label{eqn:C}
\end{equation}
where $c_{k}$ with $k=0,~1,~2$ and $3$ are the series expansion
coefficients. They are given by
\begin{equation}
 c_{0} = C(\theta_{e})~,\quad
 c_{1} = C'(\theta_{e})~,\quad
 c_{2} = \frac{1}{2}C''(\theta_{e})~,\quad
 c_{3} = \frac{1}{6}C'''(\theta_{e})~,
\end{equation}
where $C'$, $C''$, and $C'''$ are the respective first, second, and
third derivatives of $C$ with respect to $\theta$ evaluated at
$\theta_{e}$. Since the expressions of $c_{0},$ $c_{1},$ $c_{2}$ and
$c_{3}$ are too big, we have not reproduced them here.

By substituting Equations (\ref{eqn:A}) and (\ref{eqn:C}) into 
Equation (\ref{eqn:Ew2}), we obtain the components of
$\textbf{E}(\omega):$
\begin{equation} 
E_{i}(\omega)=E_{0} \int^{+\infty}_{-\infty} (a_{i0}+a_{i1} \vartheta+a_{i2}
\vartheta^{2})e^{i (c_{1} \vartheta+c_{2} \vartheta^{2}+c_{3} \vartheta^{3})} 
         d\vartheta~,
\label{eqn:Exyz1}
\end{equation} 
where $i=x,~y$ and $z.$ Here $\vartheta=\theta-\theta_{e}$ and $E_{0}= qe^{i
   [(\omega R_{0}/c)+c_{0}]}/(\sqrt{2 \pi}R_{0}c ).$

Next, by substituting the integral solutions $S_{0}$, $S_{1}$ and
$S_{2}$ given in Appendix--A, we obtain
\begin{equation}
 E_{i}(\omega)=E_{0}(a_{i0} S_{0}+a_{i1}S_{1}+a_{i2}S_{2})~.
 \label{eqn:Exyz}
\end{equation}

The Stokes parameters: $I,$ $Q,$ $U$ and $V$ have been used as tools
to specify the polarization state of the radiation field.  The net
radiation that the observer tends to receive at any rotation phase
$\phi'_{m}$ will be incoherent addition of radiation from bunches from
all those emission points which lie within the beaming
region. Following the definitions given by \citet{G10}, we estimate
the resultant Stokes parameters $I_{s},$ $Q_{s},$ $U_{s}$ and $V_{s}.$

\section{SIMULATION OF PULSE PROFILES}
\subsection{Emission From Beaming Region}
By assuming uniform source distribution and using the viewing
parameters: $\alpha=45^{\circ},$ $\sigma= 2^{\circ},$ pulsar rotation
period $P=1$~s, phase $\phi'_{m}=0^{\circ}$, normalized emission
height $r_{n}=0.01,$ particle's Lorentz factor $\gamma=400$ and
observation frequency $\nu=600$~MHz, we computed the Stokes parameters
for the radiation field. The contour plots of $I,$ $L$ and $V$ in
$(\theta,~\phi)$--plane are given in Figure
\ref{fig:br_emission_1}. For comparison we present both the cases:
non-rotating - panels ($\textit{a}$), ($\textit{b}$), and
($\textit{c}$), and rotating - panels ($\textit{a}'$),
($\textit{b}'$), and ($\textit{c}'$).  The parameters $I,$ $L$ and $V$
are normalized with the corresponding maximum value of $I$ within the
beaming region. We find the ratio of peak intensity in the rotating
case to that in non-rotating is about 2.  In all panels, the contour
levels are marked on the respective contours. Also the contours are
gray colored in such a way that darker the region lower the
corresponding parameter value.

In the non-rotating case, the trajectory of the sources are same as
their associated dipolar field lines. Hence the contours of $I$, and
$L$ are symmetric, while those of $V$ are antisymmetric with respect
to $(\theta,~\phi_{0})$ plane. In the rotating case, the contours of
$I$, $L$, and $V$ gets rotated in ($\theta,~\phi$)--plane due to
non-uniform aberration within BRP.

Using the parameters $\alpha=30^{\circ}$, $\sigma=5^{\circ}$, $r_{n}=
0.05$, $P=1$~s, $\gamma=400$ and $\nu=600$~MHz at discrete rotation
phases of the magnetic axis $\phi'_{m}=-10^{\circ},$ $0^{\circ},$ and
$10^{\circ}$, we simulated the emissions from the beaming region and
presented the contours of $V$ in Figure \ref{fig:br_emission_2}.  In
each panel $V$ is normalized with the corresponding peak intensity
$I_{peak}$ within the beaming region.  The rotation of circular
contour pattern is more towards the inner rotation phases
$(\phi'_{m}=0^{\circ})$ compared to that on outer rotation phases
$(\phi'_{m}=\pm 10^{\circ})$.  This is because, the perpendicular
distance from the pulsar spin axis to the beaming region decreases
towards the outer rotation phases, and hence the smaller rotation
effects.  Further, circular contour pattern on the trailing side
$(\phi'_{m}= 10^{\circ})$ gets more rotated compared to that on the
leading side $(\phi'_{m}=- 10^{\circ})$.  This is due to larger
asymmetry in the intrinsic curvature of the field lines between the
smaller and larger $\theta$ parts within the beaming region on the
trailing side compared to that on the leading side.

To describe the behaviors of circular polarization and the swing of
polarization position angle (PPA), we define the symbols: ``+/-'' for
transition of $V_s$ from left-handed (LH) to right-handed (RH) and
``-/+'' for transition from RH to LH.  Call the counterclockwise swing
of PPA $(d\psi_{s}/d\phi'_{m}>0)$ as ``ccw'' and the clockwise swing
($d\psi_{s}/d\phi'_{m} < 0 )$ as ``cw''.

\subsection{Emission With Uniform Source Distribution}
\begin{figure}
\centering
\epsscale{.6}
\plotone{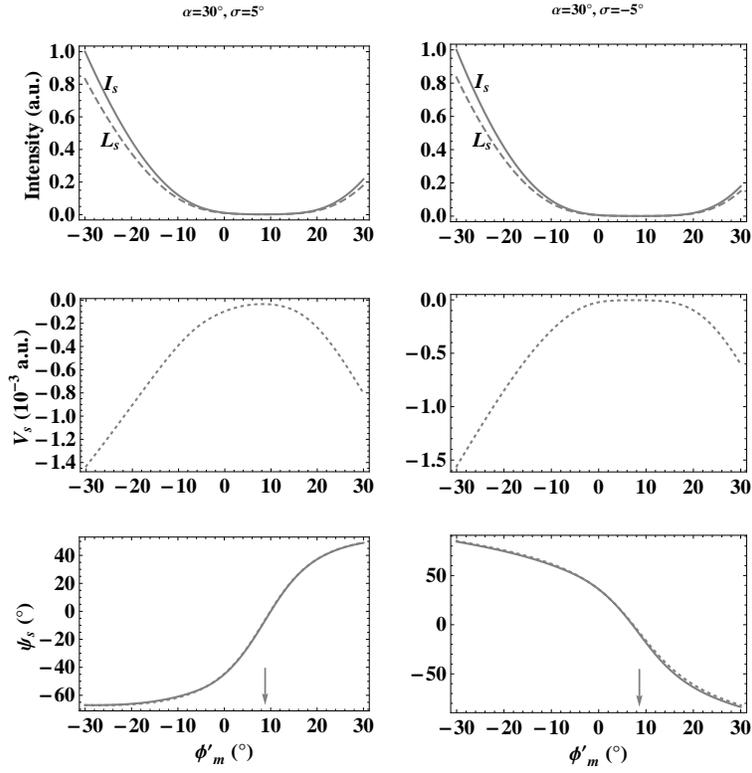}
\caption{Simulated pulse profiles: chosen $P=1$~s, $\gamma=400$,
  $\nu=600$~MHz and $r_{n}=0.05$. In the $\psi_{s}$ panels: solid
  curves are due to our simulations and the dotted curves are due to
  \citet{BCW91} model. The arrows in the $\psi_{s}$ panels mark the
  polarization angle inflection point.}
\label{fig:unmod_emission}
\end{figure}

Assuming uniform source distribution and using the viewing parameters:
$\alpha=30^{\circ},$ $\sigma=\pm 5^{\circ},$ $P=1$~s, $r_{n}=0.05,$
$\gamma=400$ and $\nu=600$~MHz, we computed the Stokes parameters for
the curvature radiation. The profiles of $I_{s},$ $L_{s},$ $V_{s}$ and
$\psi_{s}$ are plotted in Figure \ref{fig:unmod_emission}.  The
parameters in each panel are normalized with the corresponding maximum
value of $I_s.$ In both the cases of $\sigma,$ $I_{s}$ becomes more
stronger on leading side $(\phi'_{m}<0^{\circ})$ of the fiducial phase
$(\phi'_{m}=0^{\circ})$ than on the trailing side.  This is because,
due to the rotation induced curvature, the source trajectories on
leading side become more curved than on the trailing side.  The dip in
the intensity near $\phi'_{m}=0^{\circ}$ is due larger radius of
curvature $\rho$ compared to the other regions.  The behavior of
$L_{s}$ is similar to $I_{s}$ except it's smaller values due to
incoherent addition of radiation from bunches.  We observe that a
small quantity of circular polarization survives due to the rotation
induced asymmetry.  The polarization position angle is increasing
(ccw) in the case of $\sigma=5^{\circ}$, whereas it is decreasing (cw)
in the case of $\sigma=-5^{\circ}.$ The PPA inflection point, the
phase at which $|d\psi_{s}/d\phi'_{m}|$ is maximum (indicated by an
arrow), is found to be shifted to $\phi'_{m}=8.8^{\circ}$ for
$\sigma=5^{\circ}$ and to $\phi'_{m}=8.6^{\circ}$ for
$\sigma=-5^{\circ}.$ These shifts are about $3~r_{n}.$

For comparison we have superposed \citet{BCW91} PPA curves (dotted
curves) on our simulated PPA curves. Our simulated PPA profile shapes
and the shift of PPA inflection point are found to be in good
agreement with BCW (1991) model near the central parts $(\phi'_m\sim
0)$ but slightly deviated at larger rotation phases due to the
approximations made in the \citet{BCW91} model.  Also, note that at
any rotation phase $\phi'_{m}$, \citet{BCW91} model considers only the
emission from the central point of the beaming region, whereas we
consider the emissions from the whole of the beaming region.

\subsection{Emission With Modulation}
In general pulsar average radio profiles consist of many components,
which could be due to emission from plasma columns that are associated
with sparks on polar cap.  When sight line cuts through such
emissions, it encounters intensity pattern, which may be treated as
approximately Gaussians
\citep[e.g.,][]{Ketal94}.  We consider time independent modulation
\citep{G10} so that any fluctuation in the intensity strength of
individual pulses will be smoothed out and hence our simulated
profiles are expected to resemble the pulsar average profiles.
Consider a time independent Gaussian modulation in both the polar and
azimuthal directions:
\begin{equation}
f(\theta,\phi) = \sum f_0\,\exp\left[
  -(\theta-\theta_{p})^{2}/\sigma_{\theta}^{2}\right] ~ \exp\left[
  -(\phi-\phi_{p})^{2}/\sigma_{\phi}^{2}\right]~,
\label{eqn:mod}
\end{equation} 
where $(\theta_{p}, \phi_{p})$ define the peak location of the
Gaussian and $f_{0}$ is the amplitude. The parameters $\sigma_{\theta}
= \rm{w}_{\theta}/(2\sqrt{ln 2})$ and $\sigma_{\phi} =
\rm{w}_{\phi}/(2\sqrt{ln 2})$, where $\rm{w}_{\theta}$ and
$\rm{w}_{\phi}$ are the corresponding full width at half-maximum
(FWHM) of the Gaussian in the two directions. The resultant Stokes
parameters $I_{s},~Q_{s},~U_{s}~ \rm{and}~ V_{s}$ after taking into
account of modulation are given in \citet{G10}.

\subsubsection{Simulation of Core Emission}
\begin{figure}
\centering
\epsscale{.9}
\plotone{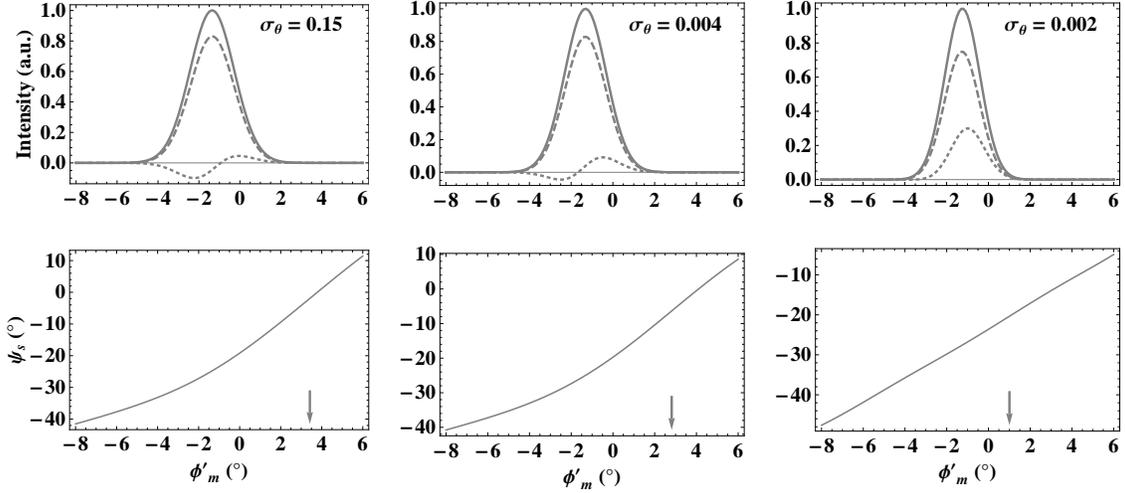}
\caption{Simulated pulse profiles: Chosen $\alpha=10^{\circ},$
  $\sigma=2^{\circ},$ $P=1$~s, $r_{n}=0.02,$ $\gamma=400$,
  $\nu=600$~MHz, $f_{0}=1,$ $\theta_{p}=1^{\circ},$
  $\phi_{p}=0^{\circ}$ and $\sigma_{\phi}=0.15$. In each panel $I_{s}$
  (solid curves), $L_{s}$ (dashed curves) and $V_{s}$ (dotted curves)
  are normalized with the respective peak intensity. The arrows in the
  $\psi_{s}$ panels mark the polarization angle inflection point.}
\label{fig:core_emission_1}
\end{figure}

To explore the effect of rotation on the central core component of
pulsar radio profiles, we consider a Gaussian modulation having peak
at ($\theta_{p},~\phi_{p})=(1^{\circ},~0^{\circ})$. We have chosen the
peak of modulation slightly away from the magnetic axis to have
modulation in both the $\theta$ and $\phi$ directions.  For
simulation, we used the viewing parameters: $\alpha=10^{\circ},$
$\sigma=2^{\circ},$ $P=1$~s, $r_{n}=0.02,$ $\gamma=400$ and
$\nu=600$~MHz.  Since the minimum of the coordinate $\theta$ is $\sim
2/3~\sigma=1.33^{\circ}$ in this case, sight line passes through the
emission region where modulation strength is slightly below its
amplitude.  To see the combined effect of rotation and modulation, we
considered $\sigma_{\phi}=0.15$ and three cases for
$\sigma_{\theta}=0.15,$ 0.004 and 0.002, and the simulated
polarization profiles are given in Figure \ref{fig:core_emission_1}.
In all the three cases of $\sigma_{\theta}$, intensity profiles are
shifted to earlier phase with respect to the fiducial phase
$\phi'_{m}=0^{\circ}$ while the polarization position angle profile
shifts to the later phase due to effect of rotation.  Note that in the
absence of rotation, the minimum of $\theta_{0}$ and the antisymmetric
point of $\phi_{0}$ (i.e., phase at which $\phi_{0}=0$) occur at
$\phi'_{m}=0^{\circ}$ and hence the intensity will peak at
$\phi'_{m}=0^{\circ}.$ But in the rotating case intensity peaks shift
to earlier phases by about $1.36^{\circ},$ $1.31^{\circ}$ and
$1.24^{\circ}$ in the three cases of $\sigma_{\theta}=0.15,$ 0.004 and
0.002, respectively.  The phase shift of $I_{s}$ peak is found to
decrease with decreasing $\sigma_{\theta}.$ The reasons are, due to
aberration, the minimum of $\theta_{0}$ and the antisymmetric point of
$\phi_{0}$ and hence the peak of modulation (phase at which observer
encounters maximum modulation strength) shift to earlier phase. But
the emission due to uniform source distribution becomes more stronger
on the leading side of the modulation peak compared to that on the
trailing side.  Therefore, the peak of modulated total intensity will
be further advanced in phase with respect to the peak of
modulation. However this extra phase shift of modulated intensity
peaks with respect to modulation peaks favor the formation of a more
symmetric component: larger unmodulated emission on smaller radius of
curvature side will be less enhanced by weaker modulation whereas
smaller unmodulated emission on larger radius of curvature side will
be more enhanced by stronger modulation. Note that if the modulation
is more steeper then the intensity profile closely follows the
modulation.  Hence the pulse width and the phase shift of peak of the
total intensity will decrease as we go from $\sigma_{\theta}=0.15$ to
$\sigma_{\theta}=0.002.$

\begin{figure}
\centering
\epsscale{.6}
\plotone{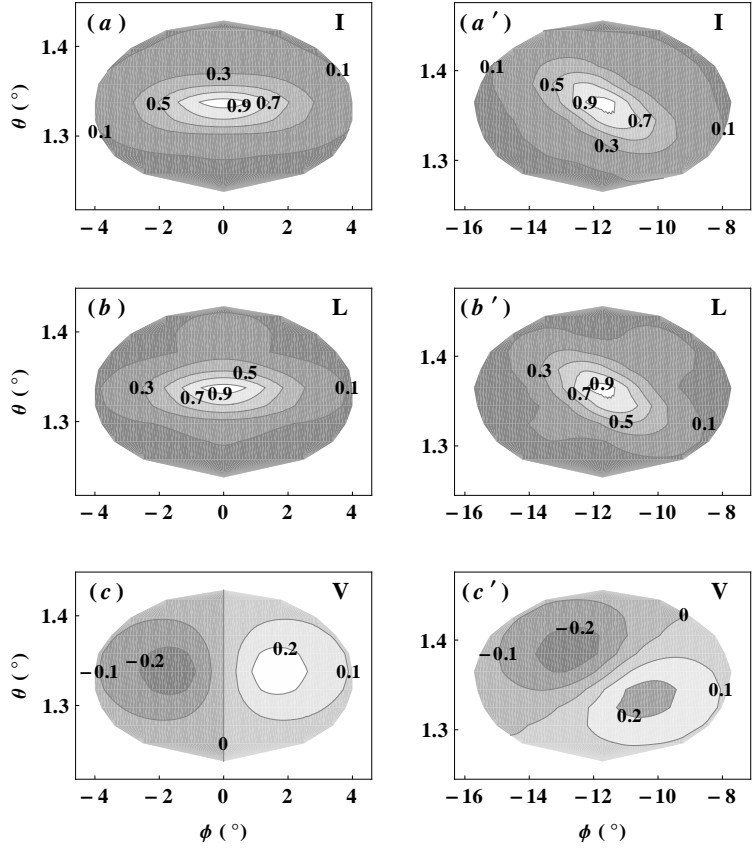}
\caption{The simulation showing the emission from a beaming
  region due to an uniform distribution of sources in the non-rotating
  (left column panels) and the rotating (right column panels) cases at
  $\phi'_{m}=0^{\circ}$. Chosen $\alpha=45^{\circ}$,
  $\sigma=2^{\circ}$, $r_{n}=0.01$, $P=1$~s, $\gamma=400$, and
  $\nu=600$~MHz.}
 \label{fig:br_emission_1}
\end{figure}

\begin{figure}
\centering
\epsscale{1}
\plotone{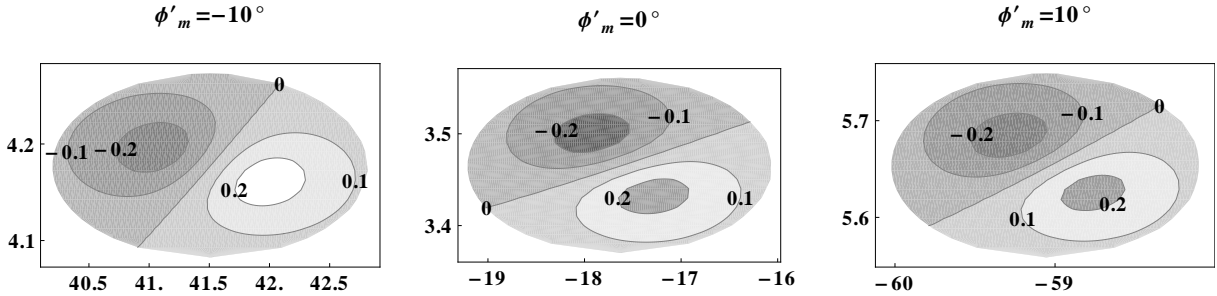}
\caption{The simulation showing the circular polarization $V$ from the
  beaming region in the rotating case at three discrete rotation
  phases. Chosen $\alpha=30^{\circ}$, $\sigma=5^{\circ}$,
  $r_{n}=0.05$, $P=1$~s, $\gamma=400$ and $\nu=600$~MHz.}
\label{fig:br_emission_2}
\end{figure}

In the case of $\sigma_{\theta}=0.15$, the circular polarization is
antisymmetric and the transition is from RH (negative) to LH
(positive).  The leading negative circular is found to be more
stronger compared to that on the trailing positive circular, whereas
they are equal in the non-rotating case \citep{G10}.  The asymmetry in
the strengths of negative and positive circulars can be explained as
follows: due to rotation , the pattern of circular polarization (see
Figure \ref{fig:br_emission_1}) gets rotated in the $(\theta,~
\phi)$--plane and hence an asymmetry is introduced in both $\theta$
and $\phi$ directions. Since we used $\theta_{p}=1^{\circ}$ and the
minimum of $\theta$ that the observer encounters is $\sim
1.3^{\circ},$ the modulation in the $\theta$ direction always enhance
the emission over smaller values of $\theta$ (see Figure
\ref{fig:br_emission_1}) compared to that over larger values of
$\theta$ within the beaming region.  Further, since we used
$\phi_{p}=0^{\circ}$, modulation in $\phi$ direction selectively
enhance the emissions over smaller values of $|\phi|$ compared to
those over larger values of $|\phi|$ within the beaming region. Since
$\sigma_{\phi}=\sigma_{\theta}$ and beaming regions are more extended
in $\phi$ compared to those in $\theta$, the modulation gradient in
$\phi$ dominates over that in $\theta$. Further, the magnitude of
rotation of circular polarization pattern within the beaming region is
more for the inner rotation phases (phases closer to
$\phi'_{m}=0^{\circ}$) as compared to that on outer phases (see Figure
\ref{fig:br_emission_2}). Hence modulation can selectively enhance the
leading negative circular over the trailing positive circular.  Also,
because of above said reasons, the phase location of the sign reversal
of circular is found to be lagging the phase location of the peak of
total intensity by a small amount.

In the case of $\sigma_{\theta}=0.004,$ the trailing positive circular
becomes more stronger than the leading negative circular, which is
opposite to $\sigma_{\theta}=0.15.$ Even though the effect of rotation
on the pattern of
circular polarization is same as in the case of
$\sigma_{\theta}=0.15,$ the modulation becomes comparatively stronger
in the $\theta$ direction as $\sigma_{\theta}\ll\sigma_{\phi}.$ Hence,
on the leading side, the net two dimensional modulation selectively
enhances emission over the lower left part of the beaming region,
whereas on the trailing side, it selectively enhance the emission over
the lower right part (see Figure \ref{fig:br_emission_2}).  Hence the
positive circular on the trailing side becomes more stronger compared
to negative circular on the leading side.  Due to above said reasons,
the phase of sign reversal of circular is found to be leading the
phase of total intensity peak by a small amount.

In the extreme case of $\sigma_{\theta}=0.002$, modulation becomes
much more stronger in the $\theta$ direction.  Therefore the
modulation selectively enhances the emission over the lower part of
the beaming region through out the pulse window.  Hence the circular
polarization becomes almost positive through out the pulse. Again due
to asymmetry in the magnitude of rotation of the circular polarization
pattern with respect to rotation phase $\phi'_{m},$ the survived
positive circular is found to be more stronger on the trailing side as
compared to that on leading side.

In all the three cases, $L_{s}$ almost follows $I_s$ except for its
magnitude.  Further, when $V_{s}$ is weaker $L_{s}$ is found to be
little stronger and vice versa.  In all the three cases, PPA swing is
`ccw' and PPA inflection points (indicated by arrows) are found to be
shifted to later phases by $3.42^{\circ},$ $2.82^{\circ}$ and
$1.00^{\circ}$, respectively.  The phase shift of the position angle
inflection point is found to decrease with decreasing
$\sigma_{\theta}$ due to the combined effect of rotation and
modulation.

\citet{BCW91} have predicted that due to aberration both the $\theta$
minimum and the antisymmetric point of $\phi$ shift to earlier phase
by $\sim r_{n},$ $(1.14^{\circ}~\rm{for}~r_{n}=0.02),$ with respect to
the fiducial phase $\phi'_{m}=0^{\circ}.$ They assumed that the
centroid of the intensity profile coincides with the $\theta$ minimum
and the antisymmetric point of $\phi$. Further, by using the particle
acceleration vector, which reflects direction of the electric field
vector in time domain, \citet{BCW91} have also predicted that the
shift of PPA inflection point to later phase by $\sim 3~r_{n},$
$(3.44^{\circ}~\rm{for}~r_{n}=0.02).$ However, in our model, we
estimate the radiation field in frequency domain. We consider the
effect of rotation along with modulation and a detailed geometry of
emission region which includes finite beaming regions from which the
observer can receive the considerable radiation, which is not
considered in the \citet{BCW91} model. Hence the phase shifts of total
intensity and PPA inflection points can be significantly different
from those predicted by \citet{BCW91}.

Note that if one considers the retardation (radiation propagation time
delay), the emissions from the beaming region at any rotation phase
$\phi'_{m}$ will be arrived at later time, i.e., by delay $\delta
t=\hat{n}\cdot\textbf{r}/c$, and hence phase delayed by $\delta
\phi'_{ret}=\Omega \delta t$ \citep[e.g.,][]{G05}. However since we
have assumed that emissions from the whole beaming region originate
from a particular altitude $r$ for a given phase $\phi'_{m}$, they
will have roughly the same $\delta t\sim r/c$ (which at most differ by
$\sim 10^{-8} s$ between center to boundary within the beaming
region). Hence emissions will arrive at the same time
$t_{r}=t_{e}+\delta t$, where $t_{e}~\rm{and}~t_{r}$ are the emission
and reception times of the radiation, respectively. Further, since we
consider a constant $r$ across the whole pulse, the net emission due
to whole beaming region at any phase $\phi'_{m}$ will be time delayed
by the same $\delta t\sim r/c.$ Hence a constant phase delay of $\delta
\phi'_{ret}\sim\Omega r/c=r/r_{LC}=r_{n}$ across the whole pulse is
introduced. After taking into account of retardation along with
aberration, the phase shifts of intensity peak and PPA inflection
point for example in the case $\sigma_{\theta}=0.15$ of Figure
\ref{fig:core_emission_1} will be
$2.50^{\circ}~\rm{and}~2.28^{\circ}$, respectively. Since retardation
just causes the shift of entire aberrated profile by $\delta
\phi'_{ret}$ to the earlier phase, we have not reproduced simulations
by combining retardation along with aberration. However the shapes of
the aberrated profile will be affected if one considers the varying
altitude across the pulse.

Also note that if one considers the modulation which is broader than
the one considered in the case of $\sigma_{\phi}=\sigma_{\theta}=0.15$
of Figure \ref{fig:core_emission_1}, then the aberration phase shift
of the intensity peak becomes substantially different from $r_{n}.$
Further $V_{s}$ becomes almost symmetric type with negative circular
through out the profile. This is because as modulation becomes broader
pulse also becomes broader.  Further in Figure
\ref{fig:core_emission_1}, we kept $\sigma_{\phi}$ constant and varied
$\sigma_{\theta}$ for the three cases. On the other hand if one
considers the case in which $\sigma_{\theta}$ is kept constant and
$\sigma_{\phi}$ is varying from more broader modulation to steeper
one, then the behavior of total intensity and PPA profiles will be
similar to Figure \ref{fig:core_emission_1}. But the evolution of
$V_{s}$ will be from almost symmetric type to antisymmetric type due
to above mentioned reasons. Also, if one considers the case in which
both the $\sigma_{\phi}$ and $\sigma_{\theta}$ vary by the same amount
from more broader modulation to steeper one, then the behavior of
pulse profiles will be similar to the case wherein $\sigma_{\theta}$
is kept constant and $\sigma_{\phi}$ is varying from larger to smaller
value. Further if one considers the negative $\sigma$ then the
polarization profiles behave similar to the positive $\sigma$ except
for the fact that the polarities of $V_{s}$ and swing of PPA profile
will be opposite.

\begin{figure}
\centering
\epsscale{.9}
\plotone{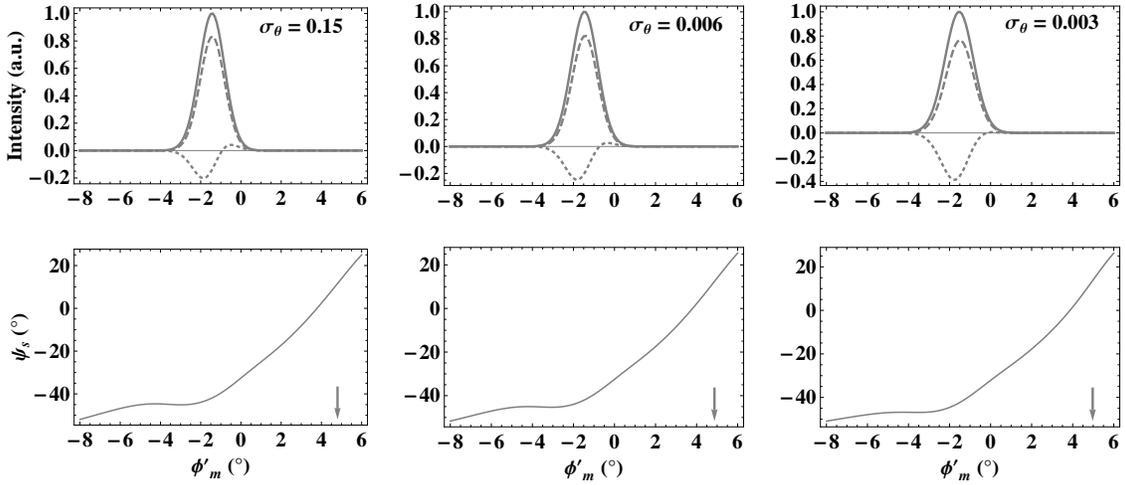}
\caption{Simulated pulse profiles: Chosen $\sigma=1^{\circ}$ and the
  other parameters same as in Figure \ref{fig:core_emission_1}.}
\label{fig:core_emission_2}
\end{figure}

To see the combined effect of rotation and modulation in the case of
sight line passing through the other side of the emission region which
lies towards the magnetic axis, we considered the sight line with
$\sigma=1^{\circ}.$ For simulations we kept the other parameters the
same as those of $\sigma=2^{\circ}$ case of
Figure~\ref{fig:core_emission_1}. The simulated polarization profiles
for $\sigma_{\phi}=0.15$ and the three cases $\sigma_{\theta}=0.15,$
0.006 and 0.003 are given in Figure~\ref{fig:core_emission_2}.  The
phase shifts of the total intensity peak in the cases
$\sigma_{\theta}=0.15,$ 0.006 and 0.003 are found to be
$1.44^{\circ},$ $1.46^{\circ}$ and $1.53^{\circ},$ respectively.  The
phase shift of the intensity peak tends to increase with decreasing
$\sigma_{\theta}$, a behavior opposite to the case of
$\sigma=2^{\circ}$ (see Figure \ref{fig:core_emission_1}).  This is
because, since we have chosen $\theta_{p}=1^{\circ}$, the emission
point coordinate $\theta$ will be closer to $\theta_{p}$ at the outer
rotation phases $\vert\phi'_{m}\vert>0^{\circ}.$ But the minimum of
the emission point coordinate $|\phi|$ will be closer to chosen
$\phi_{p}=0^{\circ}$ for the inner earlier rotation phases.
Therefore, the modulation mapped onto a broader pulse phase as
$\sigma_{\theta}$ decreases.  Hence, the phase shift of intensity peak
increases as we go from $\sigma_{\theta}=0.15$ to 0.003.

In the case of $\sigma_{\theta}=0.15$ circular polarization is
marginally antisymmetric and the transition is from negative to
positive.  The leading negative circular is found to be much more
stronger than the trailing positive circular compared to
$\sigma_{\theta}=0.15$ of Figure~\ref{fig:core_emission_1}.  Due to
viewing geometry observer tends to receive radiation from the beaming
region whose emission points in $\theta$ are always less than
$\theta_{p}=1^{\circ}$ in this case. Hence the net modulation
selectively enhances emission over the upper left part of the beaming
region (see Figure \ref{fig:br_emission_2}) on the leading side,
whereas it selectively enhances the emission over the upper right part
of the beaming region on the trailing side.  Hence from
Figure~\ref{fig:br_emission_2} we see that the leading negative
circular becomes much more stronger than the trailing positive
circular.  The phase lag of the location of sign reversal of circular
with respect to intensity peak is found to be more compared to that in
the case of $\sigma=2^{\circ}$ and $\sigma_{\theta}=0.15.$ In the case
of $\sigma_{\theta}=0.006,$ the negative circular becomes even more
stronger compared to positive circular, and in the extreme case of
$\sigma_{\theta}=0.003$, circular becomes symmetric, i.e., only
negative circular survives through out the pulse.

In all the three cases, the position angle swing is `ccw' and the
phase shifts of its inflection point for $\sigma_{\theta}=0.15,$ 0.006
and 0.003 are found to be $4.80^{\circ},$ $4.87^{\circ}$ and
$4.97^{\circ},$ respectively. The phase shift of the position angle
inflection point is found to increase with decreasing
$\sigma_{\theta}$ unlike in the case of $\sigma=2^{\circ}$ where it
decreases with decreasing $\sigma_{\theta}.$ The observed opposite
trend in the phase shift of the PPA inflection point with respect to
modulation parameter $\sigma_{\theta}$ is because of opposite trend in
the selective enhancement of the emission over a part of the beaming
region in the two cases of $\sigma$.

Note that so far we have considered the general cases of core
modulation where the modulation peak is located slightly away from the
magnetic axis and modulated in both $\theta$ and $\phi$ coordinates.
The more plausible case for core modulation is a Gaussian whose peak
is located at the magnetic axis, i.e., at $\theta_{p}=0^{\circ}$ and
depends only in $\theta.$ The corresponding modulation function that
follows from Equation~(\ref{eqn:mod}) is $f(\theta)=f_{0}\,
\exp(-\theta^{2}/\sigma_{\theta }^{2}).$ If one simulates the pulse
profiles with this modulation, $V_{s}$ becomes symmetric type:
positive circular (LH) for positive $\sigma$ and negative circular
(RH) for negative $\sigma$ through out the pulse due to the selective
enhancement. Note that in the non-rotating model \citep{G10}, the
modulation in only $\theta$ direction will never give the considerable
net circular irrespective of its gradient, as it always enhance both
the positive and negative circulars by the same amount (see panel (c)
in Figure~\ref{fig:br_emission_1}).

\subsubsection{Simulation of Conal Emission}
\begin{figure}
\centering
\epsscale{.9}
\plotone{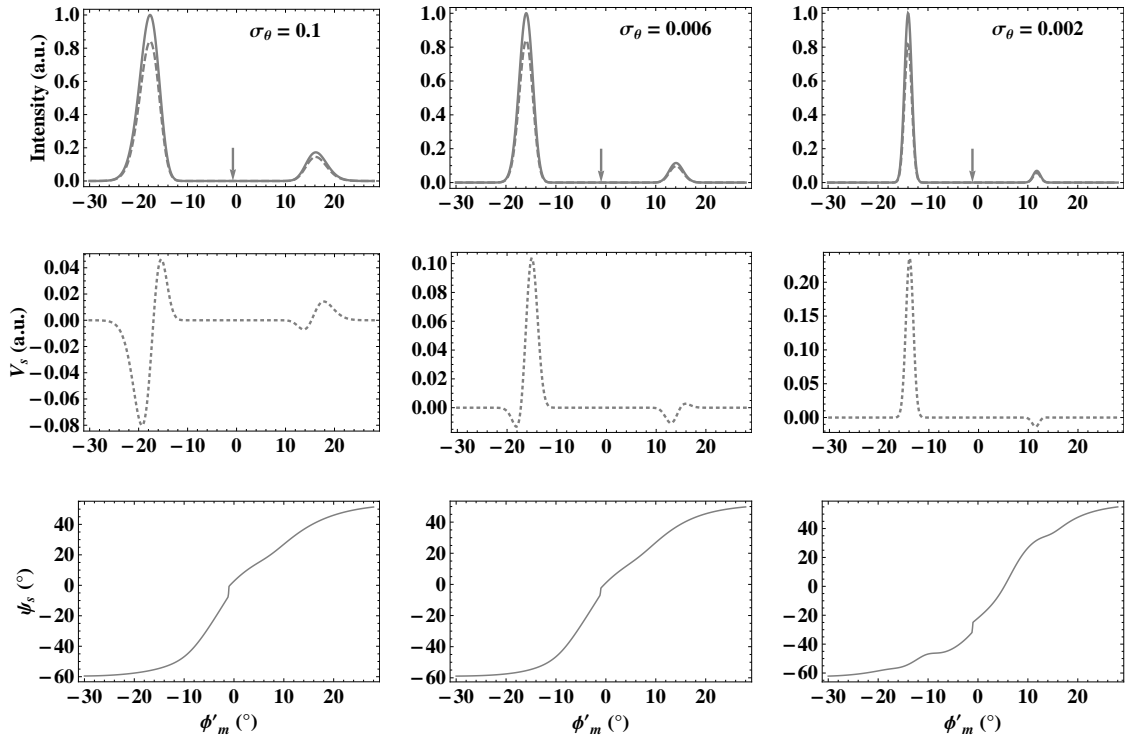}
\caption{Simulated pulse profiles: The parameters chosen are same as
  in Figure \ref{fig:core_emission_1} except 
$(\theta_{p},\,\phi_p)=(2^{\circ},\, \pm65^{\circ})$ and $\sigma_{\phi}=0.1.$}
\label{fig:cone_emission_1}
\end{figure}

In this section we consider the combined effect of rotation and
modulation on the concentric conal emissions. Consider two Gaussians
whose peaks are situated at $(\theta_{p},~\phi_{p})=(2^{\circ},~\pm
65^{\circ})$ and rest of the parameters the same as in
Figure~\ref{fig:core_emission_1}. The modulation peak locations are
chosen such that the sight line passes through the emission regions
with modulation strengths below its amplitude and the region of
maximum modulation encountered by the observer lies towards the
meridional plane. The simulated polarization profiles are given in
Figure~\ref{fig:cone_emission_1} for the three cases of
$\sigma_{\theta}=0.1,$ 0.006 and 0.002.  In all the three cases
leading side component becomes more stronger than the trailing side
component.  This is due to the combined effect of enhancement in the
intrinsic unmodulated emission and the modulation strength on leading
side over the trailing side. However the enhancement due to
unmodulated emission is more prominent. In the IOF, due to aberration,
the plasma bunch trajectories become more curved on leading side
compared to those on trailing side, and hence more emission occurs on
leading side. The marginal decrease in the strength of trailing side
component as we go from the case $\sigma_{\theta}=0.1$ to $0.002$ is
due to weaker modulation that the sight line encountered on trailing
side. Further in all the cases of $\sigma_{\theta},$ the trailing side
component becomes considerably narrower than the leading side
component. This is because, even though the modulation has roughly the
same steepness on both the leading and trailing sides, the radius of
curvature becomes more steeper in phase on trailing side compared to
that on the leading side. Hence it results in a broader component on
leading side. The phase shifts of the mid point of intensity peaks
(the cone centers indicated by arrows) to the earlier phase in the
cases $\sigma_{\theta}=0.1,~0.006,~\rm{and}~0.002$ are found to be
$0.74^{\circ}$, $0.94^{\circ}$ and $1.10^{\circ},$ respectively.

In the case of $\sigma_{\theta}=0.1,$ negative circular becomes more
stronger compared to positive circular on leading side whereas it is
vice versa on the trailing side. This is because, the magnitude of
rotation of circular pattern is more for inner rotation phases while
the net modulated emission is slightly smaller compared to outer
phases and hence selective enhancement of outer side circulars within
the intensity components. In the case of $\sigma_\theta=0.006,$
positive circular became more stronger compared to negative circular
over the leading side component and vice versa over the trailing side
component. In the case of $\sigma_{\theta}=0.1,$ the modulation is
broader and hence sight line encounters both the parts of modulation
which are lying towards and away from the meridional plane.  As
$\sigma_{\theta}$ decreases, observer tends to selectively encounter
the parts of modulation which are closer to meridional plane. This
selectively enhances the $V_{s}$ which lies towards
$\phi_{m}=0^{\circ}.$ Hence the inner sides of $V_{s}$ become more
stronger as compared to outer sides over both the leading and trailing
sides of $I_{s}.$ In the extreme case of $\sigma_{\theta}=0.002$,
$V_{s}$ becomes symmetric type, i.e., only positive on leading side
and negative on the trailing side.

In all the cases of $\sigma_{\theta}$ linear polarization $L_{s}$
profile almost follows the total intensity $I_{s},$ and PPA swing is
`ccw'. The distortions or kinks are due to combined effect of rotation
and modulation on the emissions over the beaming regions. The phase
shift of the PPA inflection point has become uncertain as the kinks
are affecting the central part of the position angle curves.  However
if one considers a case where central core component lies between two
conal components then the position angle inflection point can be found
without difficulty.

As a next case, we select the modulations with peaks at different
azimuthal locations $(\phi_p=\pm40^{\circ})$ on the same conal ring
which is considered in Figure~\ref{fig:cone_emission_1}. In this case
sight line passes through the emission regions with modulation
strengths below its amplitude and the region of maximum modulation
encountered by the observer lies away from the meridional plane. By
keeping other viewing parameters same as in the case of
Figure~\ref{fig:cone_emission_1}, we simulated the polarization
profiles for the cases of $\sigma_{\theta}=0.1,$ 0.006 and 0.002 and
given in Figure~\ref{fig:cone_emission_2}.  The small increase in the
strength of the trailing side component as we go from
$\sigma_{\theta}=0.1$ to 0.002 is due to an increase in the modulation
strength that the observer encounters on the trailing side. The phase
shift of the cone centers to the earlier phase in the cases
$\sigma_{\theta}=0.1,$ 0.006 and 0.002 are found to be $1.07^{\circ},$
$1.06^{\circ}$ and $1.08^{\circ}$, respectively.  These shifts are
found to be almost independent of $\sigma_{\theta}$ unlike in the
previous cases where they considerably dependent on the
$\sigma_{\theta}.$ In the case of $\sigma_{\theta}=0.1,$ $V_{s}$ is
antisymmetric on both the leading and trailing sides with the outer
circular is more stronger than the inner.  This is similar to the
cases $\sigma_{\theta}=0.1$ in Figure~\ref{fig:cone_emission_1}, where
the modulation is broader.  In the case of $\sigma_{\theta}=0.006$,
$V_{s}$ is again antisymmetric on both leading and trailing sides but
outer circular dominates over the inner.  This is because, sight line
encounters the major part of the modulation, which lies away from the
meridional plane.  Hence there is a selective enhancement of the outer
circular relative to the inner. In the extreme case of
$\sigma_{\theta}=0.002$, $V_{s}$ becomes symmetric on both the sides:
negative on leading side and vice versa on the trailing. These are of
opposite behaviors as compared to Figure~\ref{fig:cone_emission_1}. In
all the three cases of $\sigma_{\theta}$, $L_{s}$ almost follows the
$I_{s}$ with lower values similar to the previous cases, and the PPA
swing is `ccw' and shows the kinky behavior.

\section{DISCUSSION}
Pulsar rotation along with modulation and viewing geometry seems to be
greatly influencing the pulsar radio profiles.  Due to rotation the
trajectories of sources on leading side becomes more curved compared
to those on trailing side, and hence leading side unmodulated emission
always dominate over the trailing one.  If one considers an
azimuthally symmetric cone modulations then the leading side intensity
components become more stronger than the corresponding trailing ones
due to the rotationally induced asymmetry in the curvature of plasma
trajectories \citep[e.g.,][]{TG07}. In support of this there is a
strong observational result by \cite{LM88}, and our simulations
clearly confirm it. We also find that the leading side intensity
components become wider than the corresponding trailing ones due to an
asymmetry in the gradient of radius of curvature between leading and
trailing sides.  These findings have an observational evidence
\citep{AG02} and a similar behavior has been discussed by
\citet{DWD10}.  Note that if we consider higher emission altitude in
our simulations then the trailing side component gets substantially
weaker or even vanishing.  Hence it could serve as an explanation for
the ``partial cones'' \citep{Tetal10}.

The fact that the phase shift of the intensity components to earlier
rotation phases and that of the PPA inflection point to later phase is
a natural consequence of effect of rotation.  The phase shifts of the
centroid of pulse and that of the PPA inflection points have been
predicted to be about $r_{n}$ and $3 r_{n}$, respectively
\citep{BCW91}.  But in their simplistic model considered only the
emission from the points at which source velocity vector exactly
aligns with the observer's sight line.  However there is a
considerable emission from the other points of beaming region, which
is influenced by rotation and modulation.  As a result the shifts of
intensity component and PPA inflection point will no longer remain as
$r_n$ and $3 r_n,$ respectively.  We have shown that, due to pulsar
corotation, the pattern of emission within the beaming region gets
rotated in $(\theta,\phi)$--plane, and hence an asymmetry is
introduced in both the $\theta$ and $\phi$ directions.  Due to these
asymmetries within the beaming region, either the antisymmetric or
symmetric type circular polarization become possible depending upon
the viewing geometry and modulation.  If the modulation is more
steeper and has roughly the same gradient in both $\theta$ and $\phi$,
then the antisymmetric circular polarization is observed.  On the
other hand the symmetric type circular polarization is more plausible
when modulation is broader and has roughly the same gradient in both
$\theta$ and $\phi,$ and also when modulation is more steeper in
$\theta$ than in $\phi.$ But in literature, circular polarization has
been modeled only in the non-rotating pulsar approximation. Hence only
the antisymmetric circular polarization was thought to be a natural
feature of curvature radiation, and the symmetric type circular
polarization was speculated to be a consequence of  propagation
effect \citep[e.g.,][]{Getal93,G10}.

\begin{figure}
\centering
\epsscale{.9}
\plotone{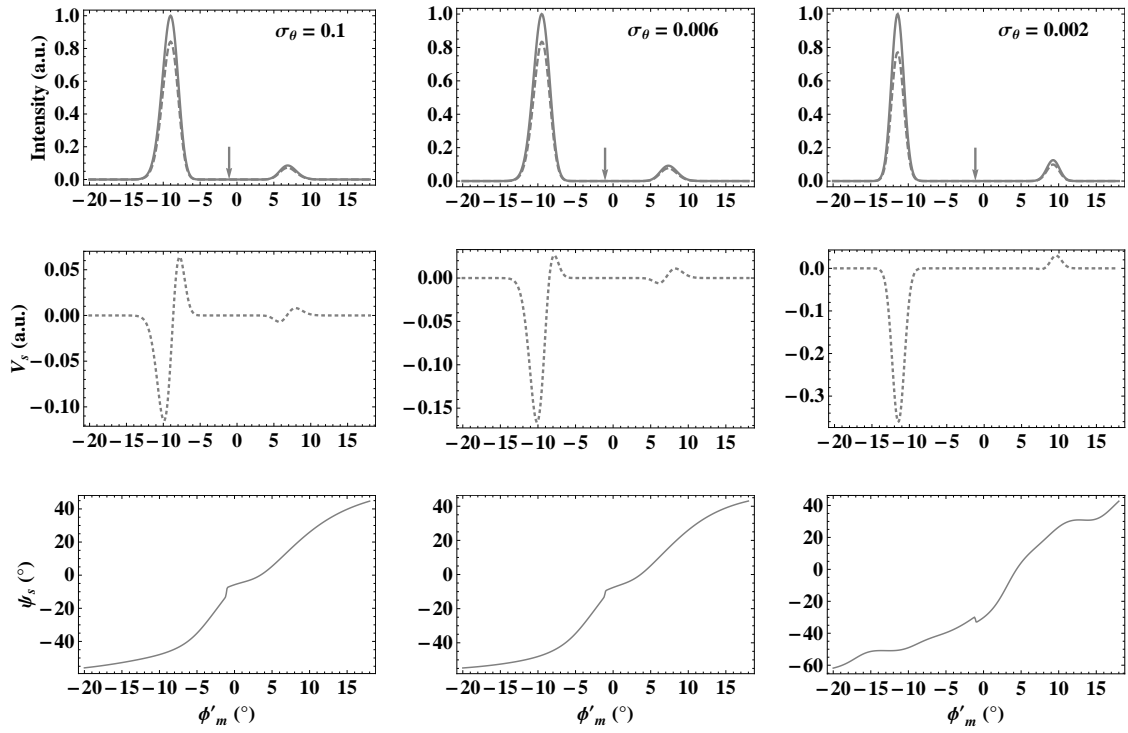}
\caption{Simulated pulse profiles: The parameters chosen are same as
  in Figure~\ref{fig:cone_emission_1} except
$\phi_p=\pm40^{\circ}.$}
\label{fig:cone_emission_2}
\end{figure}

\citet{Hanetal98} and \citet{YH06} have found that the sign reversal
of circular polarization is not only associated with the central
`core' region but also found over conal components as well as at the
intersection of conal components.  From our simulations of conal
components, it is possible to explain all types of circular
polarization sense reversals and its association with either
increasing or decreasing PPA of Table~3 in \citet{YH06}.  For example,
consider the Figures \ref{fig:cone_emission_1} and
\ref{fig:cone_emission_2} in the case of $\sigma_{\theta}=0.1.$ On the
leading side, we get a case where the circular polarization changes
sign from negative to positive with an increasing PPA.  Again by
considering the $\sigma_{\theta}=0.002$ cases of Figures
\ref{fig:cone_emission_1} and \ref{fig:cone_emission_2}, on leading
side we can get a case where circular polarization changes sign from
positive to negative with increasing PPA.

We do confirm the Radhakrishnan \& Rankin's (1990) correlation that
the sense reversal of circular polarization from negative to positive
is correlated with `ccw' PPA swing (or increasing PPA) and vice versa,
and we argue this as a geometric property of curvature radiation.  In
pulsars with `symmetric' type of circular polarization,
\citet{Hanetal98} have not found any correlation between the sense of
circular polarization and the PPA swing.  Our simulations also
indicate that the negative circular polarization can be associated with
either `ccw' or `cw' PPA swing depending upon the viewing geometry and
modulation locations. Similarly the positive circular too can be
associated with `ccw' or `cw' PPA swing.  However, \citet{Hanetal98}
and \citet{YH06} have found that many conal--double pulsars show a
single handed circular polarization over both the components. The
negative circular correlated with the `ccw' PPA swing and the positive
circular with the `cw' PPA swing.

Our simulations of conal components show that if the sight line is
missing the modulation peaks and the steepness of modulation in the
polar ($\theta$) direction is much larger as compared to that in the
azimuthal ($\phi$) direction, then the circular polarization becomes
single handed on both the leading and trailing sides but have opposite
signs (see for example case $\sigma_{\theta}=0.002$ of Figures
\ref{fig:cone_emission_1} and \ref{fig:cone_emission_2}).  Note that
we considered situations where the leading and trailing side
modulations symmetrically lie on a cone centered on the magnetic axis.
On the other hand if one considers a situation where the modulations
are asymmetrically located on a cone, then the correlation between the
sense of circular polarization and the PPA swing in the case of
conal-double pulsars can be explained.  For example, by choosing the
locations of modulations in the case of $\sigma_{\theta}=0.002$ of
Figure~\ref{fig:cone_emission_2} at $\phi_{p}=40^{\circ}$ and
$-65^{\circ}$, one can get negative circulars over both the leading
and trailing components (see the cases $\sigma_{\theta}=0.002$ of
Figures \ref{fig:cone_emission_1} and \ref{fig:cone_emission_2}), and
hence an association of negative circular with the increasing PPA can
be established.

The `kinky' type distortion in PPA profile has been found in some
normal pulsars and more commonly in millisecond pulsars.
\citet{Metal00} have attributed this effect to multi polar magnetic
field while \citet{MS04} have speculated that the
aberration/retardation resulting from the height-dependent emission
can cause the distorted PPA traverses.  \citet{RK03} following
\citet{HA01} have proposed that the discrete jumps in the PPA profiles
are due to magnetospheric return currents.  However, from our
simulations it is clear that even with a constant emission altitude
across the whole pulse profile the `kinky' behaviors can be produced
in PPA traverses. Due to an incoherent addition of radiation field
emitted from a beaming region, which is affected by rotation and modulation,
the distortions in the PPA traverses are introduced.
The PPA traverse under both the core as well as conal components are
found to get distorted. But there are observational claims that
central core region is more likely to show RVM distortions than the
conal regions \citep[e.g.,][]{R83,R90,RR90}.

In this work we considered a constant emission altitude $r$ across the
whole pulse to make comparison with the earlier results. However
varying emission altitude across the pulse can be incorporated. We
considered only the rotation and time independent modulation effects
in our simulations, as we are interested in the polarization properties
of curvature radiation which is of intrinsic origin. We plan to consider
the propagation effects, polar cap currents, magnetic field sweep back
and higher multi polar components of magnetic field on pulsar radio
emission in our future works.

\section{CONCLUSION}

By developing a relativistic model for pulsar radio emission we have
attempted to explain the complete polarization state of the curvature
radiation.  Our model takes into account of a very detailed geometry
of emission region, rotation and modulation as detailed in
section~3.3, which have not been much incorporated in the earlier
models.  Based on our pulse profile simulations, we conclude the
following:
\begin{enumerate}
\item The phase shift of intensity components to earlier phase and the
  PPA inflection point to later phase, are strongly influenced by the
  combined effect of rotation and the modulation.
\item The components on the leading side become stronger and
  broader than those on the trailing side because of
  rotation.
\item In an unmodulated emission a small quantity of circular
  polarization survives due to rotationally induced asymmetry, but
  from the point of view of observations it is insignificant.
\item For the very first time we are able to show that the `symmetric'
  type circular polarization can be obtained within the frame work of
  curvature radiation. This result is very important from the point of
  view of emission mechanism.
\item Both the types of circular polarization: antisymmetric 
   $(+/-$ or $-/+)$ and symmetric
  $(+$ or $-),$ can result any where within the pulse window
  due to the combined effect of rotation, viewing geometry and
  modulation. This might be responsible for the diverse nature of 
  circular polarization.
\item We argue that pulsar rotation combined with modulation can 
  introduce `kinky' patterns into the PPA traverses.
\end{enumerate}
\begin{acknowledgements}
We thank J. L. Han and Pengfei Wang for stimulating discussions, and
anonymous referee for useful comments.
\end{acknowledgements}
\vfill\eject
\renewcommand{\theequation}{A-\arabic{equation}}
\setcounter{equation}{0}  
\section*{APPENDIX--A:}
Consider the integral
\begin{equation}
S_{0} = \int\limits_{-\infty}^{+\infty} {e}^{i(c_{1}\, \vartheta+c_{2}\, 
      \vartheta^2+c_{3}\,\vartheta^3}) d\vartheta~.
\end{equation}
By changing the variable of integration $\vartheta = (x/l)+m,$ and defining
the constants $l = \sqrt[3]{3c_3}$ and $m = -c_2/(3c_3),$ we obtain
\begin{equation}
\int\limits_{-\infty}^{+\infty} {e}^{i(c_{1}\,\vartheta+c_{2} \,
        \vartheta^2+c_{3}\, \vartheta^3)}d\vartheta 
      = U \int\limits_{-\infty}^{+\infty} e^{i \left(z x +\frac{x^3}{3}\right)}dx~, 
\label{I0}
\end{equation}
where
$z = \frac{1}{\sqrt[3]{3c_3}}\left(c_1-\frac{c_2^2}{3 c_3}\right)$ and 
$U = \frac{1}{\sqrt[3]{3c_3}} e^{i \frac{c_2}{3 c_3}\left(\frac{2 c_2^2}
    {9 c_3}-c_1\right)}. $

For ${\rm Im}(z) = 0,$ we know
\begin{equation}
j_{0} = \int\limits_{-\infty}^{\infty}{\rm e}^{i\left(z x+\frac{x^3}{3}\right)} dx 
    = 2 \pi Ai(z),
\end{equation}
where Ai$(z)$ is an entire Airy function of z with no branch cut 
discontinuities, and
\begin{equation}
j_{1} = \int\limits_{-\infty}^{\infty} x{e}^{i\left(z x+\frac{x^3}{3}\right)}dx
    = -i 2 \pi Ai'(z)~,
\end{equation}
where $Ai'(z)$ is the derivative of the Airy function $Ai(z).$
Therefore, we have
\begin{equation}
S_{0} = U~j_{0}~.
\end{equation}
By differentiating equation~(\ref{I0})  with respect to $c_1,$
we obtain
\begin{eqnarray}
S_{1} = \int\limits_{-\infty}^{+\infty}\vartheta {e}^{i(c_{1} \vartheta+c_{2} 
        \vartheta^2+c_{\rm 3} \vartheta^3)}d\vartheta
  & = & \frac{U}{\sqrt[3]{3c_3}}\int\limits_{-\infty}^{+\infty}
        \left(x-\frac{c_2}{\sqrt[3]{3c_3^2}}\right) 
        e^{i \left(z\, x +\frac{x^3}{3}\right)}dx~\nonumber \\
  & = &\frac{U}{\sqrt[3]{3c_{3}}}\left(j_{1}-\frac{c_{2}}
        {\sqrt[3]{9c_{3}^2}}j_0\right) ~.
\end{eqnarray}
Differentiation of equation~(\ref{I0}) with respect to
$c_2$ gives
\begin{eqnarray}
S_{2} = \int\limits_{-\infty}^{+\infty}\vartheta^2 {e}^{i(c_{1}\vartheta+c_{2}
          \vartheta^2+c_{3} \,\vartheta^3)}d\vartheta
  & = & \frac{U}{3c_3} \int\limits_{-\infty}^{+\infty}
        \left(\frac{2 c_2^2}{3c_3}-
        c_1-\frac{2c_2}{\sqrt[3]{3c_3}}x\right) 
        e^{i \left(z x +\frac{x^3}{3}\right)}dx~\nonumber \\
  & = & \frac{U}{3c_3}\left[\left(\frac{2 c_2^2}{3c_3}-c_1\right)j_{0}-
        \frac{2c_2}{\sqrt[3]{3c_3}}j_{1}\right]~. 
\end{eqnarray}

\clearpage 

\end{document}